\documentclass[11pt, NumberedRefs]{article}
\usepackage{setspace} 
\usepackage{multirow}

\usepackage{amsmath}
\usepackage{amsfonts}
\usepackage{graphicx}

\usepackage{url}
\usepackage{fullpage}

\usepackage{tabularx}
\usepackage{authblk}

\begin{document}
\title{Seabed Classification using Physics-based Modeling and Machine Learning}

\author[1]{Christina Frederick}
\affil[1]{Department of Mathematical Sciences,  New Jersey Institute of Technology, Newark, NJ 07102, U.S.A.}
\author[2]{Soledad Villar}
%\email{soledad.villar@nyu.edu}
\affil[2]{Center for Data Science and Courant Institute of Mathematical Sciences, New York University, New York, NY 10011}
\author[1]{Zoi-Heleni Michalopoulou}
%\email{michalop@njit.edu}
%\affiliation{Department of Mathematical Sciences,  New Jersey Institute of Technology, Newark, NJ 07102, U.S.A.}

\date{\today}

\maketitle
\begin{abstract}
In this work model-based methods are employed along with machine learning techniques to classify sediments in oceanic environments based on the geoacoustic properties of a two-layer seabed.  Two different scenarios are investigated.  First, a simple low-frequency case is set up, where the acoustic field is modeled with normal modes.  Four different hypotheses are made for seafloor sediment possibilities and these are explored using both various machine learning techniques and a simple matched-field approach.  For most noise levels, the latter has an inferior performance to the machine learning methods.  Second, the high-frequency model of the scattering from a rough, two-layer seafloor is considered. Again, four different sediment possibilities are classified with machine learning. For higher accuracy, 1D Convolutional Neural Networks (CNNs) are employed. In both cases we see that the machine learning methods, both in simple and more complex formulations, lead to effective sediment characterization. Our results assess the robustness to noise and model misspecification of different classifiers.
	\end{abstract}
	
\section{Introduction}\label{sec:forward}

	Sonar classification of the ocean floor relies on the quality of the acoustic images and the data-processing tools used for interpretation. The pixel resolution in sonar images depends on a number of factors, including the instrument range, geometry, and operating frequency. Low-frequency sonar instruments towed far from the ocean floor cover large swath-widths with lower resolution capabilities. Higher frequency sonars that are towed closer to the ocean floor have less coverage but can produce higher pixel resolution.  For example, GLORIA operates at about 6 kHz frequency and has a daily coverage of around 20,000 km${^2}$, producing images with $
	\sim$60 meter pixel resolution. TOBI operates at 30 kHz frequency, has a daily coverage of 470 km${^2}$, and produces images with $\sim$10 meter pixel resolution. The typical pixel resolution is then about 20$\lambda$, where $\lambda$ is the acoustic wavelength.
	
	Since the measured acoustic response from the ocean floor is produced by multiple sophisticated scattering processes, there is no simple way to interpret each pixel of a sonar image in terms of seafloor parameters such as sediment type, roughness, and layer structure.  Well-known approaches such as matched-field processing \cite{Baggeroer1988a, Tolstoy2000}, image-processing based techniques \cite{Rzhanov2012, Reed2006, Cervenka1993}, and recently, machine learning algorithms, are successful in known environments for which ground-truth datasets are available \cite{Gorman1988,Gorman1988a, Perry2001}. 
	
	To increase the performance and applicability of current machine learning algorithms, the main challenge is overcoming the lack of large training datasets. In this paper, we demonstrate the potential of physics-based forward modeling to generate training data for seafloor classification problems. The advantages of physics-based modeling for these problems include the ability to accurately label data and fully explore the possible seafloor parameter space. This enhances the training of machine learning models, as well as the evaluation of the reliability and validity of the outcomes.

	 There are two main contributions of our work aimed at better understanding the limits of modern techniques applied to scenarios with complex physics and limited observational data. The first is physics-based training, in which training datasets are designed using numerical simulations of representative segments of the seafloor domain. At lower frequencies, these template domains are characterized by sound speed, attenuation, and layer thickness. At higher frequencies, finite element modeling (FEM) can incorporate additional small-scale features such as the roughness of the water-sediment interface. The second contribution is a set of machine learning experiments for classifying the top sediment in a two-layer seafloor. In the low frequency case, these results are compared with known techniques, and conclude that machine learning tools provide higher classification accuracy. At high frequencies, deep learning offers a significant improvement in performance over shallow machine learning tools. To our knowledge, this is the first work that attempts material classification from synthetic backscatter generated with FEM-based simulations.

%scattering from rough, layered seafloors. Motivated by recent work on classification of material type and roughness parameters in a single layer seabed \cite{Engquist2017},
\section{Inverse wave problems in ocean acoustics}
\label{sec:inverse}

\begin{figure}
	\includegraphics[height=0.4\textwidth]{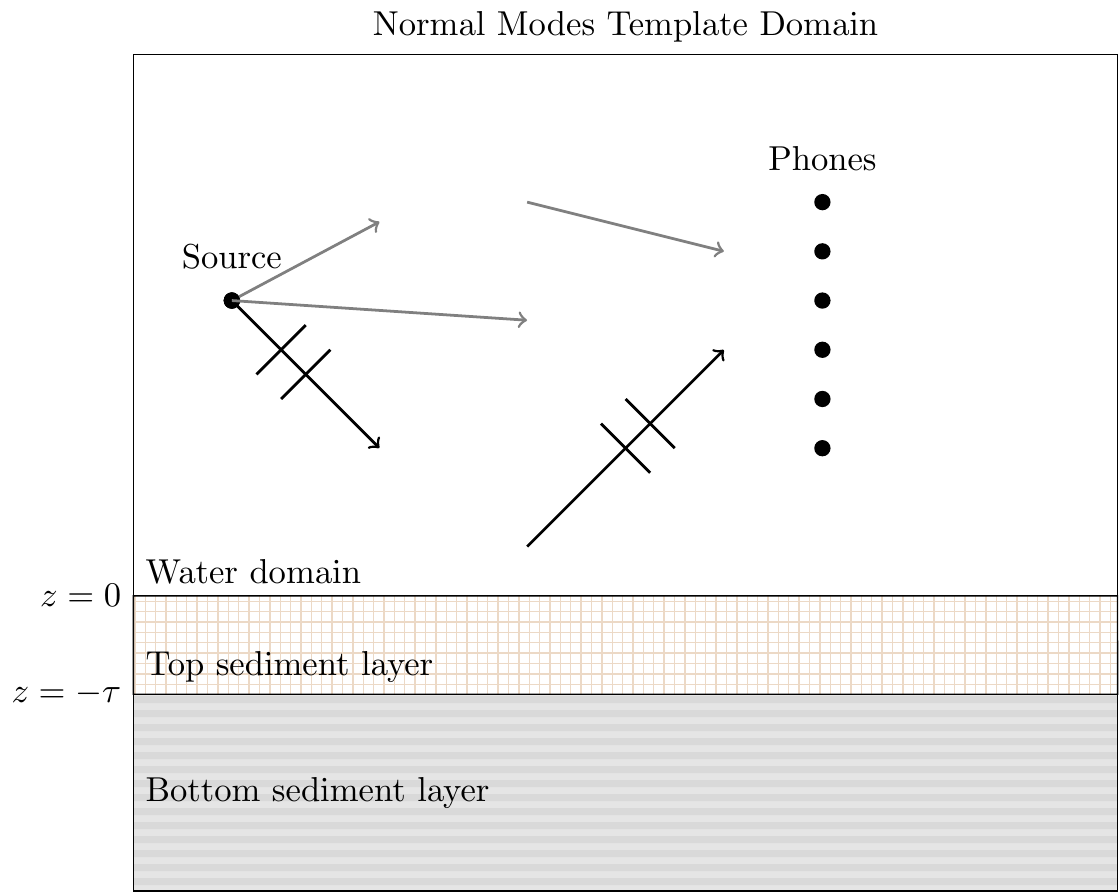} \includegraphics[height=0.4\textwidth]{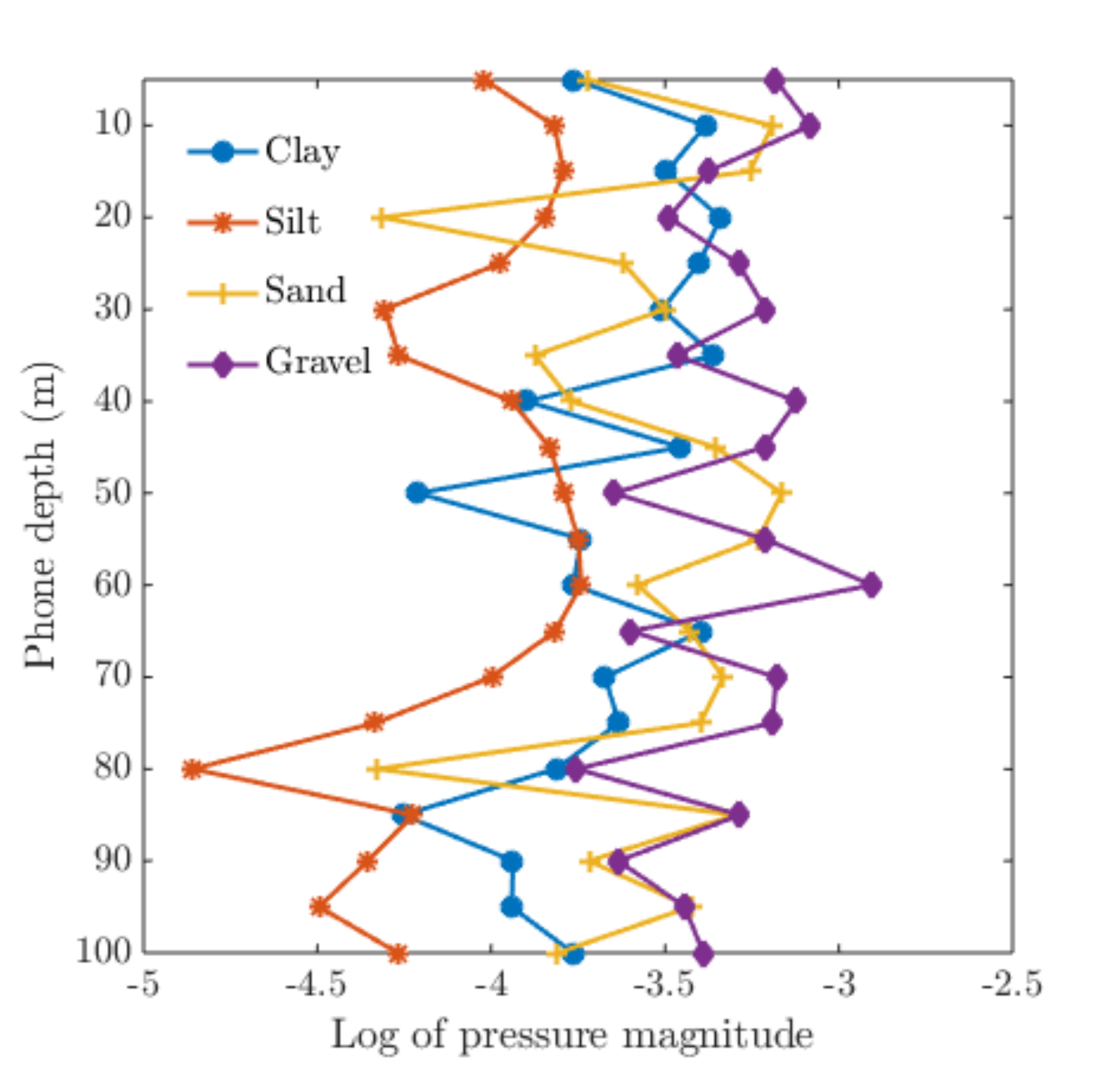}
	\caption{(left) Sound propagation in an oceanic waveguide. (right) Simulated pressure field measurements for four top sediments.}\label{fig:krakensetting}
\end{figure}

For the present work, we model the ocean as a range-independent waveguide with homogeneous water and two sediment layers  (see Figures \ref{fig:krakensetting} and \ref{fig:fwd}). As in the majority of underwater acoustic modeling \cite{Chapman2001, Jackson2007,  Jackson1986}, we treat the sediment layer, as a fluid, and model time-harmonic acoustic pressure waves at range $x$ and depth $z$ as solutions $p(x,z)$ to the Helmholtz equation, 
\begin{eqnarray}
%{\rho_{0}}\nabla\cdot\left(\frac{1}{\rho_{0}} \nabla p\right)  +\frac{\omega^{2}}{c_{0}^{2}} p =0 \text{ in }  D_{w}, \label{eq:fwd1}\\
%p \vert_{+} = p\vert_{-},  \frac{1}{\rho_{0}}\frac{\partial p}{\partial \vec{n}}\vert_{+} =\frac{1}{\rho}\frac{\partial p}{\partial \vec{n}}\vert_{-}  \text{ on }  z=f(x).\label{eq:fwd2}\\
{\rho}\nabla\cdot\left(\frac{1}{\rho} \nabla p\right)  +\frac{\omega^{2}}{c^{2}} p +i\alpha {\omega}{p} =s,\label{eq:Helmholtz}
\end{eqnarray}
where $\omega$ represents the angular frequency, and the spatially varying functions $\rho$, $c$, and $\alpha$, represent the density,  sound speed, and attenuation, respectively, of the medium (in seawater, attenuation is negligible at low and intermediate frequencies and it is reasonable to set $\alpha=0$). The source $s$ and boundary conditions depend on the instrument and environment, and coupling conditions on the water-sediment and sediment-sediment interfaces enforce continuity of the pressure and normal displacement along the fluid-sediment and sediment-sediment interfaces.

The mathematical formulation of the inverse problem is to determine a vector of true parameters $m^{*}\in\mathcal{M}$ using a finite set of  observables ${\bf Y}$. Here,  $\mathcal{M}$ is a space of unknown parameters, for example, source location, bathymetry, water column sound speed, and geoacoustic and geometric properties of the sediment layer (density, compressional and shear wave speeds, etc.).  In sonar imaging, the data is a set of pressure or backscatter measurements ${\bf Y} = \{p(x_{k}, z_{k})\}_{k=1}^{n}$ recorded at $n$ locations $(x_{k}, z_{k})$. The  forward operator $\mathcal{F}:\mathcal{M}\rightarrow \mathbb{R}^{n}$ describes the wave propagation process relating the observables to the model parameters.

The inverse problem of determining $m^{*}$ given ${\bf Y} =\mathcal{F}(m^{*})$ is usually difficult and ill-posed; a number of different seafloor environments can produce similar measurements and that similarity can be further complicated in the presence of noise. While full waveform inversion (FWI), a standard tool in seismic imaging, achieves success by matching the data with simulations of the foward model, there is no universal technique for solving (\ref{eq:Helmholtz}) at frequencies of interest in sonar imaging. Factors such as the source-receiver geometry, frequency and bandwidth, and boundary conditions must be taken into account in formulating a solution technique. Often, a combination of analytic and numerical methods are implemented based on the environmental complexity, desired resolution, and availability of labeled ground truth data. Here we focus on two solution techniques, normal mode propagation in simple geometries and finite element modeling of the scattering process in a more complicated environment having a rough water-sediment interface.

\subsection{Background on geoacoustic inversion and sediment characterization}
\label{sec:geoacoustic}

Knowledge of the propagation medium is of paramount importance in source detection, localization, and identification in the ocean. Geoacoustic inversion is an area of research that allows us to obtain knowledge about the ocean sediments with which the propagating sound has interacted to facilitate the tasks mentioned above.
Extensive research and significant advances have been made through the years in estimating sediment parameters using the full field or select features.

Matched-field inversion (MFI) is a popular approach for inverting for (estimating) geoacoustic properties using measurements of the acoustic field at a number of receiving phones.
It stems from matched field processing (MFP)~\cite{Tolstoy1993}, originally designed for source localization, which was extended to invert for environmental parameters with the first effort presented in~\cite{Livingston1989}.  Estimates are obtained by comparing measured acoustic fields to replica-predicted fields computed with sound propagation models for numerous values of the unknown parameters.  The values that provide the best match between replicas and data are considered to be the desired estimates.  MFI initially required significant computation entailing searches in multi-dimensional domains.  Global optimization approaches have facilitated the inversion making the process less onerous~\cite{Collins1992, gerstoft1994inversion, Shorey1994, Dosso2002, Knobles2003, Michalopoulou2004}.

Backscattering strength data have also been used for sediment characterization~\cite{Michalopoulou1994, Michalopoulou1996, caiti1996acoustic, Steininger2013, Frederick2017, Zou2019} with a variety of methods using backscatter models and statistical techniques.  These approaches facilitated inversion both for seafloor roughness and sediment geoacoustic properties.

\subsection{Machine learning for seafloor classification} \label{sec:related_work}

	In the late '90s and early 2000s, feed-forward neural networks gained attention in the field of ocean acoustics because of the ability to overcome issues faced by traditional methods, such as error estimation and time-consuming global or local searches in the parameter space \cite{stephan1998neural, Chakraborty2003,michalopoulou1995application}. In these problem settings,  neural networks provided efficent, nonlinear approximators of inverse functions. 
	
	One drawback of neural networks (and many modern deep learning approaches) is the large amount of training data needed to ensure accurate, reliable performance.  Some of the earlier works overcome this using synthetic training data and real test data \cite{benson2000geoacoustic}. 
	Over the last 15 years, feature-based machine learning techniques have also been developed for problems such as target detection and recognition, source localization, and seabed classification \cite{Coiras2007,Reed2003, kong2019machine}. While these methods can be fully trained with smaller data sets, they require a manual, often tedious, feature extraction process. 

Convolutional neural networks (CNNs) offer an alternative approach in which features are instead learned by the network. The neural network can be viewed as a composition of functions, called layers, and the training process consists of determining weights in each layer using batches of labeled data and stochastic optimization.  Different layers are used for different purposes; for example, convolutional layers contain filters that are convolved with small patches of the previous layer to capture local features, and max pooling layers downsample the input by taking the maximum over a patch. Layers in a CNN are connected via an entry-wise nonlinear activation function. The most common activation function is $\text{ReLU}(x)=max(x,0)$. Modern CNN architectures can contain many layers and connections between them that form acylic graphs. The complexity and depth of the architecture can capture features and interdependencies of high-dimensional data on a wide range of scales.

 The use of CNNs in acoustics is now a very active and rich area of research \cite{Bianco2019}. Although some studies can be performed using smaller training data sets, \cite{Williams2020, Ye2019, Berthold2018,Vankomen2019b}, most deep learning approaches must develop techniques for accomodating the lack of real sonar data. This can be done by, for example, using a pretrained CNN (trained using optical images) \cite{Kvasic2019, zhu2017deep}, applying image processing techniques for data augmentation \cite{Coiras2007,Xu2017, Ding2016, Nguyen2020}, or training with synthetic data generated by a physics-based forward solver \cite{benson2000geoacoustic, niu2017source, ren2012waveguide}. Large sonar images can also be decomposed into many ``patches" \cite{Berthold2018, Scheide2018, Chen2019} that form training datasets. Our modeling, described next, is based on a combination these patch-based techniques for data and physical modeling.

\section{Seafloor characterization with pressure fields at vertical line arrays}

\subsection{Problem setup}

We consider the scenario of Figure~\ref{fig:krakensetting}, where a source transmits a continuous wave (CW) signal at frequency $\omega=2\pi f$.  A point source located at $x_{s}=0$ and depth $z_s$, determines the source term $s(x, z)=-\frac{\delta(x)\delta(z-z_s)}{2\pi x}$.
The pressure field at a depth $z$ and a distance $x$ from the source can be seen as the solution of the Helmholtz equation~(\ref{eq:Helmholtz}). Here $\rho(x, z)\equiv \rho_{0}$, that is, there is no depth dependence.  The environment is considered to be range independent and the sediment is treated as a fluid.  Solving the Helmholtz equation corresponds to $\mathcal{F}$ of Section~\ref{sec:inverse}.

Assuming that source and receiver location, bottom depth, and water column sound speed are known, we can invert for sediment sound speed $c$ and attenuation $\alpha$ via MFI (for the experiments that we consider the field is not sensitive to density, for which we cannot reliably invert).  An inner product between the normalized pressure field (solution of Eq.~\ref{eq:Helmholtz}) computed for different $c$ and $\alpha$ values and the received acoustic field can be calculated.  Those values that maximize the inner product are the parameter estimates.

In this work, we are interested in classification rather than parameter estimation.  As will be discussed later, MFI can still be employed 
towards this task, but, here, we are mostly interested in investigating the potential of machine learning in sediment classification. Recently, a sensitivity analysis was performed on parameters in a two layer seafloor, indicating the promise of these methods \cite{Neilsen2018}.
We investigate a low frequency case with the sound being transmitted by a source and received at a vertical line array with 20 hydrophones 
(see Figure~\ref{fig:krakensetting}).  The phone spacing is 5 m and the source frequency is 400 Hz.  
The source-array distance is 10 km and the source depth is 50 m.  The water column sound speed profile $c(z)$, where $z$ is depth,  
is a typical shallow water downward refracting profile and the water depth is 111 m.  One isovelocity sediment layer is assumed with 
varying sound speed $c$ and attenuation $\alpha$, representing different sediment types (clay, silt, sand, gravel) over a chalk halfspace.  
The sediment thickness is $\tau=5$ m.

\subsection{The data}

Using normal modes~\cite{Porter2017} we synthesize the pressure field at the VLA for the four sediment types; each field has 20 components corresponding to the 20 VLA receivers; these are equispaced with the first one located at a depth of 5 m.  Values of 1500, 1575, 1650, and 1800 m/s are the assumed sound speeds for clay, silt, sand, and gravel, 
respectively~\cite{Jensenbook}.  Values of 0.2, 1, 0.8, and 0.6 dB$/\lambda$, where $\lambda$ is wavelength, are the corresponding attenuations.  These sediment sound speed-attenuation values are listed in the first column of Table~\ref{table:datasets}.  One-thousand noisy field realizations are then synthesized for each sediment type by adding zero mean complex Gaussian noise.
% with a variance corresponding to Signal to Noise Ratios (SNR) between xx and yy dB.
The generated noisy fields are used to train a set of classifiers.
The noise-free fields for each class are shown in Figure~\ref{fig:krakensetting}.
%\begin{figure}
%	\includegraphics[width=\linewidth]{fig_log_pressure}
%	%\includegraphics{fig_log_pressure}
%\caption{Pressure fields for clay, silt, sand, and gravel.}\label{fig_log_pressure}	
%\end{figure}
%These are employed as testing patterns with the results shown in …
Subsequently, we perturb the sediment sound speed and attenuation values.  The new $c$ and $\alpha$ values still represent the same sediment types but now with variation. We create environments for ten sets for each sound speed-attenuation pair 
(listed in Table~\ref{table:datasets_nm}) and one-thousand noisy realizations for each case for a total of ten-thousand pressure fields. Then, the responses are the observables ${\bf Y}$ of Section~\ref{sec:inverse} that form the training and test sets.  %Data were generated for the same SNRs that were used for the training pattern generation.

Data generation for both training and testing sets was repeated for a number of Signal-to-Noise-Ratios (SNRs), because we were interested in identifying if the relative classifier performance was noise dependent.

%In each experiment, a training set is formed using labeled acoustic signals  $\{({\bf X}_{j}, y_{j})\}_{j=1}^{N}$ generated for varied environmental parameters. The labels correspond to one of four material classes based on the sound speed of the top sediment layer, $y_{j} = c_{top}$. Each classifier is learned during a training stage from training data, and the performance of the classifier is evaluated on a separate test set.

\begin{table}[h]
\footnotesize{
		\begin{tabular}{lcccccccccccc}
		
			Parameter  & Training & Test 1 & Test 2 & Test 3& Test  4 & Test 5 & Test 6 & Test 7& Test  8 & Test 9& Test  10  \\
			\hline
			
			$c$ (m/s)  &&&&&&&&&&\\
			Clay      & 1500 & 1517 & 1521 & 1517 & 1546 & 1517 & 1529 & 1526&1518 & 1527 & 1518  \\
			Silt 	& 1575 & 1577 & 1592 & 1582 & 1574 & 1591 & 1586 & 1596 & 1581 & 1584 & 1580  \\
			Sand 	& 1650 &1658 & 1632 & 1663 & 1647 & 1652 & 1644 & 1642 & 1672 & 1648 & 1655  \\
			Gravel  & 1800 & 1794 & 1795 & 1799 & 1796 & 1792 & 1802 & 1809 & 1791 & 1801 & 1784 \\
			\hline
			$\alpha$ (dB/$\lambda$) &&&&\\
			Clay& 0.2 &0.271 & 0.077 & 0.296 & 0.175 & 0.224 & 0.160 & 0.066 & 0.256 & 0.215 & 0.166  \\
			Silt  & 1.0 &1.050 & 1.064 & 0.954 & 0.892 & 0.982 & 1.039 & 1.188 & 1.046 & 0.922 & 1.189\\
			Sand & 0.8 & 1.042 & 0.843 & 0.914 & 0.708 & 0.839 & 0.681 & 0.970 & 0.881 & 0.595 & 0.797 \\
			Gravel& 0.6& 0.495 & 0.683 & 0.547 & 0.458 & 0.666 & 0.616 & 0.740 & 0.651 & 0.680 & 0.775\\
			\hline
			$\tau$ (m) & &&&\\
			All & 5 & $\ast$ & $\ast$ & $\ast$& $\ast$& $\ast$& $\ast$& $\ast$& $\ast$& $\ast$& $\ast$\\	
			
		\end{tabular}
		}
	\caption{Environmental parameters varied to generate synthetic data with KRAKEN. An '$\ast$' indicates the parameters used are the same 
		as those used in the training data.}\label{table:datasets_nm}
\end{table}

\begin{table}[h]
\begin{center}
\footnotesize{
		\begin{tabular}{lcccccc}
			Parameter & Units & Training & Test 1 & Test 2 & Test 3& Test  4\\
			%&  & $N=80000$ & $N=5000$& $N=5000$ & $N=5000$& $N=5000$\\
			%	\hline
			\hline
			Roughness &&&&&\\
			RMS height & (cm) & .5&.46&.46&.46&.46\\
			RMS corr. len. & (cm) &2& 1.5&1.5&1.5&1.5\\
			\hline
			Sound Speed &&&&&\\
			$(c_{top}, c_{bottom})$&&&&&\\
			Clay & (m/s)& 	(1500, 5250) & $\ast$ & (1501.57, $\ast$) & $\ast$ & (1501.57, 5254.58) \\
			Silt & (m/s)	&(1575, 5250)& $\ast$ & (1577.03, $\ast$) & $\ast$  & (1577.03, 5254.65) \\
			Sand & (m/s)	& (1650, 5250)& $\ast$& (1648.13, $\ast$)&  $\ast$&    (1648.13, 5246.58)  \\
			Gravel & (m/s) &(1800, 5250)& $\ast$ & (1802.07, $\ast$) &   $\ast$& (1802.07 ,5254.71) \\
			\hline
			Density \hfill  &&&&&\\
			$(\rho_{top}, \rho_{bottom})$\hfill  &&&&&\\
			Clay&(kg/m$^{3}$)&(1500, 2700)& $\ast$ & (1500.66, $\ast$) & $\ast$& (1500.66, 2704.57)\\
			Silt &(kg/m$^{3}$) & (1700, 2700)&$\ast$  & (1697.99, $\ast$) & $\ast$&(1697.99, 2699.85)\\
			Sand &(kg/m$^{3}$)& (1900, 2700)& $\ast$ & (1898.89, $\ast$)&$\ast$ & (1898.89, 2703.00) \\
			Gravel&(kg/m$^{3}$)&(2000, 2700)& $\ast$ & (1802.07, $\ast$)& $\ast$ & (1802.07, 2696.42) \\
			\hline
			Thickness&&&&&$\{0.253, 0.504,$&\\
			$\tau$ &(m) & $\{.25, .5, .75, 1\}$& $\ast$ &$\ast$&$ 0.725, 1.004\}$&$\ast$\\
		\end{tabular}
}
\end{center}
	\caption{Environmental parameters varied to generate synthetic backscatter data. An '$\ast$' indicates the parameters used are the same as those used in the training data. Among the parameters that are fixed throughout all of the data are the acoustic frequency $\omega=15$kHz, incident angle $\theta=\pi/12$ radians, and domain size of $2\times 2$ meters. }\label{table:datasets}
\end{table}

\subsection{Sediment Classification} \label{sec:baseline}

\begin{figure}[h]
\begin{center}
	\includegraphics[height=0.34\textwidth]{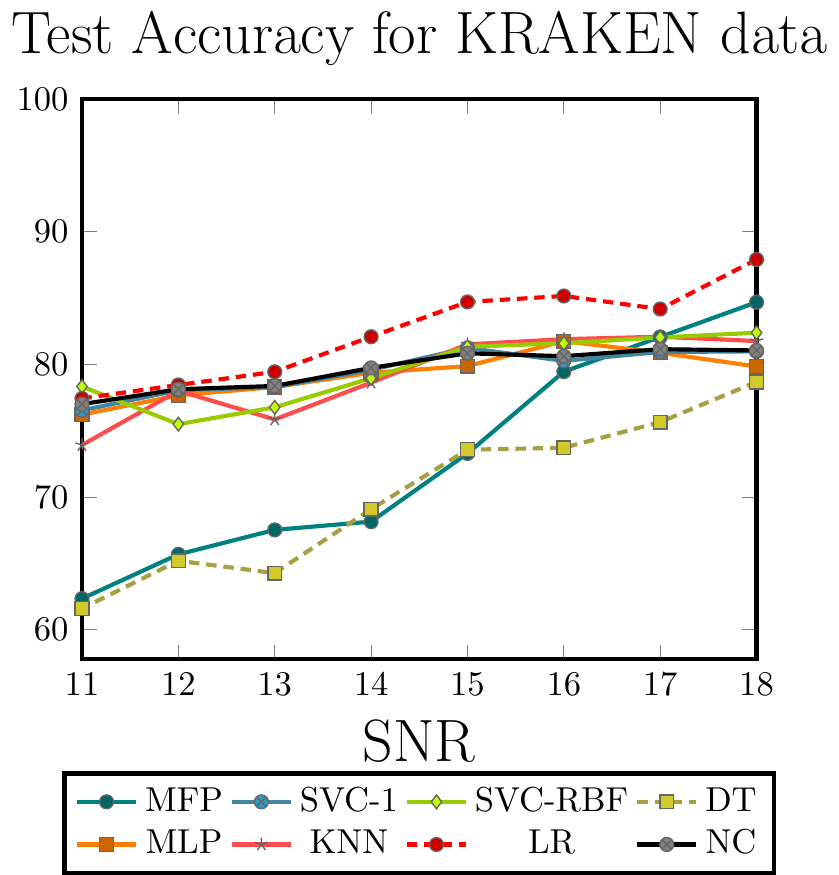} \includegraphics[height=0.34\textwidth]{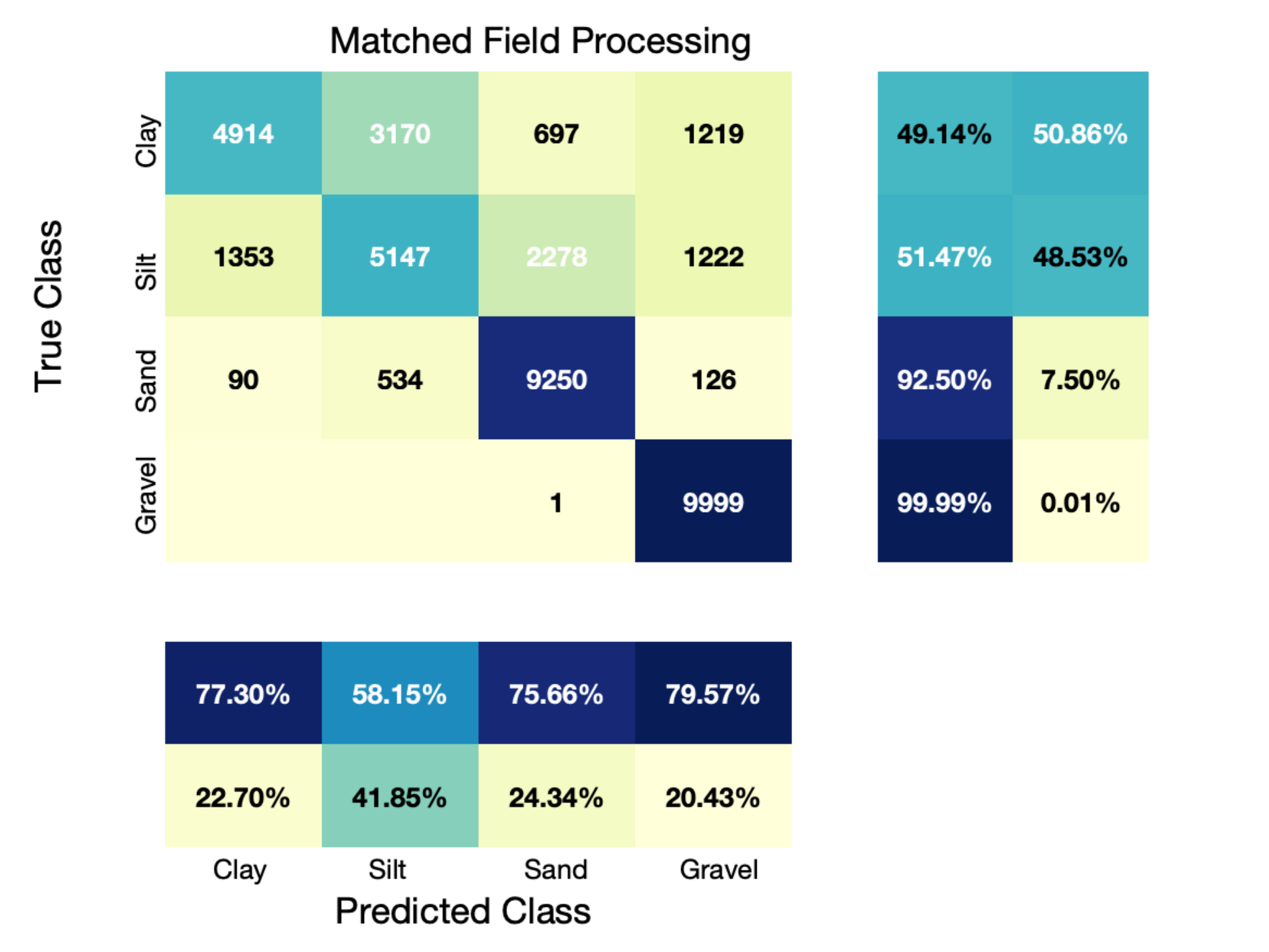}  
	
	\caption{(left) Classification performance of different trained machine learning classifiers on test data. Confusion matrices for (middle) MFI and (right) LR on noisy KRAKEN data (SNR = 18 dB).}
	\label{fig.KrakenResults}
	\end{center}
\end{figure}

The goal of our experiments is to determine whether sediment class can be determined correctly when the classifiers are tested on data 
generated for environments that are similar but not identical to those used for the generation of the training data.

Before we implemented machine learning techniques, we constructed a classifier based on MFI; that served well for the performance evaluation of the proposed techniques.  The MFI/MFP classifier relied on the computation of a simple inner product between the data to be classified and normalized replica fields calculated for the nominal values of sediment properties.  The replica fields for the four classes are the measurements shown in Figure~\ref{fig:krakensetting}.
Specifically, inner product $P$ is evaluated as:
\begin{equation}
P(m)=|{\bf w}(m)^*{\bf d}|,
\label{eq:mfp}
\end{equation}
where ${\bf w}=\frac{\bf p}{||{\bf p}||}$.  Vector ${\bf p}=[p_1 \ldots p_L]^T$, where $L=20$, consists of the solutions of the Helmholtz equation at  receivers $i=1,\ldots,20$ and $m=(c,\alpha)$.  Vector ${\bf p}$ consists of the noisy pressure measurements, that is, observables ${\bf Y}$, but calculated for the perturbed parameters.
The classifier decided the class by identifying which of the four inner products $P$ had the highest value.  The MFP classifier required no training.

The following baseline classifiers were implemented in \texttt{scikit-learn} \cite{scikit-learn}: grouping methods, such as $k$-nearest neighbors (KNN) and Nearest Centroid (NC), Support Vector Classifiers that find the best separating hyperplane (SVC-1) or kernel (SVC-R), ensemble methods such as Decision Tree Classifiers (DT) and Random Forests (RF), feed-forward neural nets (MLP), and Logistic Regression (LR). The hyperparameters of each classifier were determined using a randomized grid search over a range of possible parameters. Five-fold cross validation is used for model selection.

The training data set consisted of the 4,000 labeled noisy pressure field measurements for the nominal values of the four sediment classes. We tested the methods on ten sets of 1,000 realizations, each set for perturbed pairs of sediment parameters.  That is, no testing signal was generated for the exact $c$ and $\alpha$ values used to generate the training set vectors. Figure \ref{fig.KrakenResults} shows the overall accuracy of different classifiers. For high noise levels (low SNR), MFP seems to perform well but LR classification exhibits the best performance. While linear regression models fit data with a linear function, logistic regression models instead use the more flexible sigmoid function that is less sensitive to outliers. Unlike linear classifiers, LR models can produce nonlinear decision boundaries, which may improve the ability to discriminate between points that are close together but belong to different classes.

As the signal-to-noise ratio (SNR) increases, there is a marked difference between MFP and other processors, which all outperform MFP.  Figure \ref{fig.KrakenResults} shows the confusion matrix for MFP and the support vector machine for an SNR of 15 dB. We observe that MFP confuses silt with gravel, whereas the LR incorrectly classifies about half of the clay instances as silt, but has a very good performance 
in the rest of the materials. An interesting inference can be made from the Logistic Regression proccess results.  If a sediment is classified as clay, sand, or gravel, it is fairly likely that the characterization is correct.  In the case of silt, however, there is a possibility of misclassification. Identifying a feature that further distinguishes clay from silt measurements could potentially resolve the ambiguity.

\section{Backscatter}
\begin{figure}
	\includegraphics[width=0.6\textwidth]{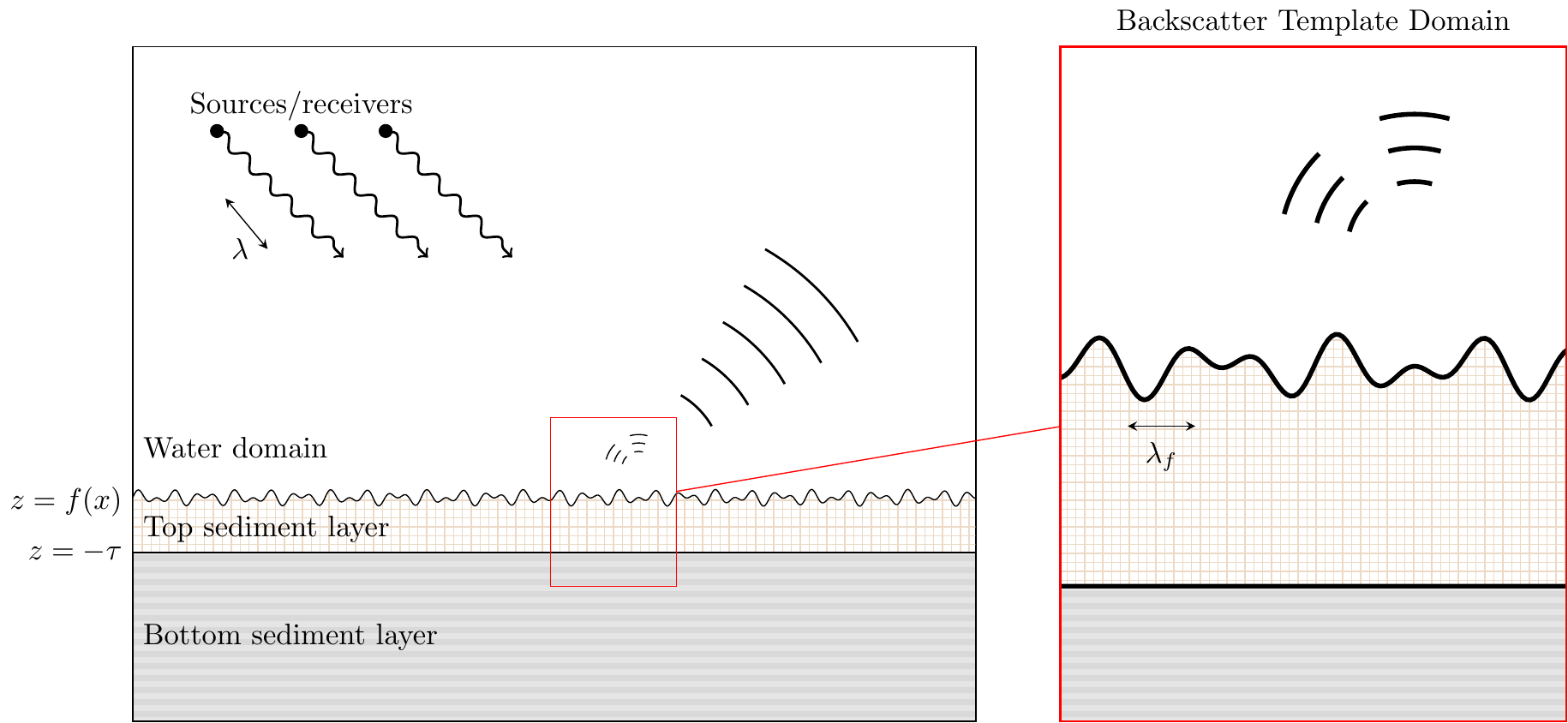} \hspace{15pt}
	\includegraphics[width=0.33\textwidth]{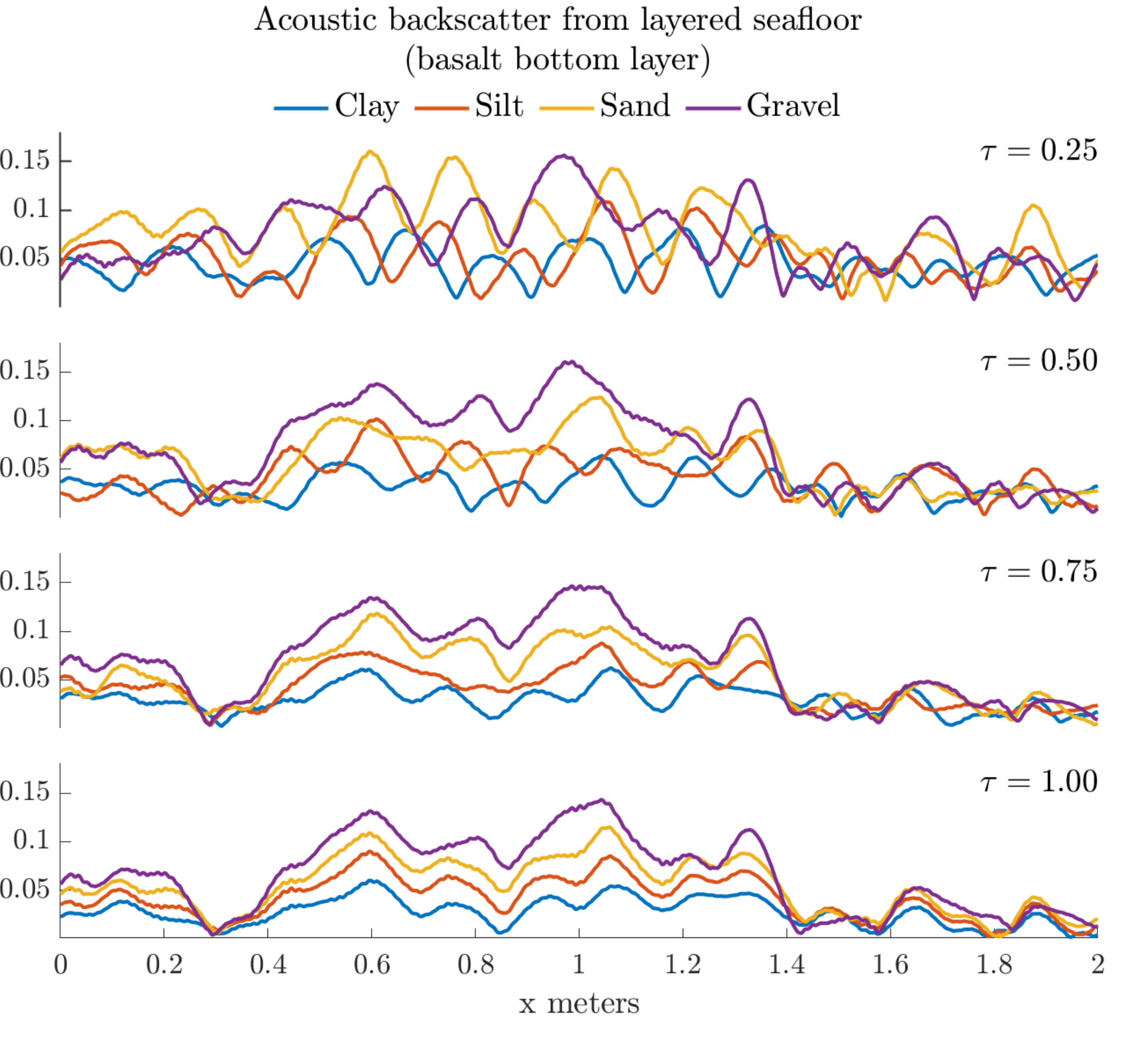}	\caption{(left) High-frequency scattering from a rough, two-layer environment. (right) Approximate backscatter from segments of the seafloor containing the water-seafloor interface.}\label{fig:fwd}
\end{figure}

On one hand, if the spatial scale of the seafloor roughness is much smaller than the acoustic wavelength, a source ping smoothly reflects from the seafloor, and analytical methods described in the previous section are preferable because of the accuracy and low computational cost. On the other hand, when the spatial scale of seafloor heterogeneities is similar to the acoustic wavelength, these methods fail to approximate the measured backscatter, and more accurate modeling is needed.

Finding exact solutions to (\ref{eq:Helmholtz}) can be very difficult, even for homogeneous environments with simple geometries, and at higher frequencies, the problem becomes even more complex, with features such as roughness complicating the task of modeling and classification. In the case of rough, layered sediments, numerical methods such as the finite element method (FEM) provide an attractive option. The FEM discretization of the governing acoustic wave equations in underwater domains enables accurate modeling of the seafloor environment  \cite{Isakson2014,Bao1995, Bruno1991, Shirron2006, Vendhan2010, Isakson2008, Jensen2010, Huynha}. Not limited by constraints on the degree of surface roughness or range dependence, FEM offers flexibility and can incorporate physics generated from realistic scenarios. In computational ocean acoustics applications, FEM is usually used for providing benchmark solutions, short range scattering, or as a part of a hybrid approach \cite{Kargl2014, Murphy2011, Abawi1997, Trofimov2015} in which the scattering is approximated using a FEM discretization of a small region near the seafloor and the reverberant field is modeled using traditional, more efficient techniques.

The primary disadvantage of FEM and other numerical methods for solving high-frequency Helmholtz equations is the high computational cost of approximating $O(\lambda^{-1})$ unknowns in each spatial dimension, where $\lambda$ is the acoustic wavelength. Computing solutions for large spatial domains in the high-frequency regime, in which each forward solve may need to resolve up to billions of unknowns, is a notoriously hard problem and a topic of intense interest in the scientific computing community \cite{Engquist2018}. In practice, the balance between accuracy and computational cost often results in model simplifications that limit the recovery of small-scale features in a range of complex underwater environments. A main challenge in ocean acoustics is to find ways to improve existing models in a way that is both physically justified and computationally efficient.

In order to address scalability of the high-frequency models to large seafloor domains, we divide the region near the seafloor into segments and perform local simulations on these subdomains. The resulting simulations produce acoustic templates from small, representative seafloor environments that can be assumed to have homogeneous properties. Our rationale for this model reduction is based on the framework of multiscale modeling for partial differential equations, in which a low-cost {\it large scale} solver of an effective model is coupled with a {\it fine scale} solver that resolves detail. The range of validity for these methods is problem-dependent and our justification is based on physical assumptions of scale-separation and spatial stationarity in the problem.

The forward process can be summarized in terms of three main stages: (a) the incoming wave from the source to the sea floor; (b) scattering and reflection from the seafloor; and (c) traveling waves recorded by receivers (See Figure \ref{fig:fwd}). Stages (a) and (c) are relatively easy to model for a wide range of source frequencies. It should be noted that variable wave propagation velocity in the water can easily be handled by existing tools, e.g. geometrical optics. The critical interactions with the seafloor that occur in stage (b) deserve accurate modeling at high frequencies to resolve the scattering effects occur near the ocean bottom that contribute significantly to the measured signal. In this case, fine scale features such as seafloor roughness must be incorporated.

Given the availability of tools for modeling wave propagation at lower frequencies, we therefore limit the present discussion to modeling the high-frequency scattering that occurs in stage (b) and testing the performance of these models in parameter inversion. Though it is out of the scope of the present work, it is an important problem to further analyze multiscale couplings between the different stages in the forward process and find ways to systematically build hybrid algorithms that adapt to more complicated physical processes occuring in the ocean.

Below we will discuss how we approach the backscatter modeling for high frequency scattering and what efforts we are making towards sediment characterization.

\subsection{Problem setup}

We consider models for acoustic backscatter on template domains shown in Figure \ref{fig:fwd} (left). When the acoustic wave generated by a source reaches this template domain, it can be modeled as an incident plane wave traveling in the direction given by the vector $k_{\theta}=\frac{\omega}{c_{w}}(\sin \theta, -\cos \theta)$, where $\theta$ is the incident angle with respect to the $z$ axis and $c_{w}$ is the sound speed in water, \begin{align}
p_{in}(x,z) = e^{-i  k_{\theta} \cdot (x,z)}.\label{eq:pinc}
\end{align}

The interaction of these waves with the seafloor produces scattered waves in multiple directions due to objects on the seafloor, the roughness of the water-sediment interface caused by sand ripples, as well as heterogeneities in the sediment, including interactions between multiple sediment layers. The total pressure field $p$ solves the homogeneous equation (\ref{eq:Helmholtz}) with $s=0$ and can be expressed in the water domain as the sum of the incident field and the scattered wavefield, $p=p_{in}+p_{scatt}$.

Sufficiently far from caustics, $p_{scatt}$ can be approximated by the superposition
\begin{align}
p_{scatt}(x,z)\simeq \sum_{\beta}P_{\beta}(x,z) e^{ i k_{\beta}\cdot (x,z)},\label{eq:wavedec}
\end{align}
where $P_{\beta}(x,z)$ is the complex amplitude of the outgoing wave traveling in the direction $\beta$ and is assumed to be locally smooth and independent of wavelength. For colocated sources and receivers, the backscattered direction is $\tilde{\theta}=\pi/2+ \theta$, and we model the received signal as backscatter measurements, \begin{align}
{\bf Y}= \left(|P_{\tilde{\theta}}(x_{k}, z_{*})|\right)_{1\leq k\leq n} \label{eq:backscattered}.
\end{align}
where $\{(x_{k}, z_{0})\}$ is a set of observation points located at a fixed depth $z\equiv z_{0}$ above the seafloor.

For unlayered seabeds, there has been significant progress in quantifying the dominant contributions to backscatter at high frequencies \cite{Broschat2002, Fuks2000, Jaud2012, Kuperman1989, Jackson2016}. Many approaches rely on cancellation effects from neighboring waves to simplify the model, and the situation is even more complicated when there are multiple sediment layers within the acoustic penetration range. Numerical approaches for approximating far-field scattering from targets using FEM include boundary-element methods \cite{Shirron2006, Gerstoft2005} and techniques for post-processing solutions using the Helmholtz-Kirchoff (HK) boundary integral formulation of the scattering problem \cite{Zampolli2008}. Another technique called numerical microlocal analysis (NMLA) \cite{Benamou2004, Landa2011} aims to isolate direction angles and amplitudes of waves crossing an observation location $(x_{0},z_0)$ without relying on knowledge of the solution on the entire boundary of a scatterer. The main idea is that if the acoustic wavelength $\lambda=2\pi/k$ contains variations on a scale sufficiently small with respect to the geometry, the solution $p(x,z)$ to $(\ref{eq:Helmholtz})$ behaves like a finite superposition of plane waves (\ref{eq:wavedec}). Then, discrete measurements of the solution on a observation circle, $p(r\cos(t_{j}), r\sin(t_{j}))$, can be expressed as a 2D Jacobi-Anger expansion that contains the amplitudes and angles of interest. By filtering the vector of observations using the fast Fourier transform, the method recovers the dominant directions and amplitudes of scattered waves. Recently, research and implementations for high-frequency Helmholtz equations further improve the accuracy and stability of the method \cite{Benamou2011, Fang2018}. 

The plots in Figure \ref{fig:fwd} (right) show the acoustic response approximated by FEM and NMLA at $z_{0}=1.5$ from a rough, two-layer seafloor model. In particular, the left plot in Figure \ref{fig:fwd} shows approximate backscatter signals $|P_{\tilde{\theta}}(x, z_{0})|$ in (\ref{eq:backscattered}) modeled with FEM and NMLA on the domain $0<x<2, -2<z<2$ for environments with varying environmental parameters $(c, \rho , \tau)$ and a fixed realization of the water-sediment interface. The signals corresponding to the thickest top layer, $\tau=1$ m,  are visually well-separated owing to the high attenuation of the signal in the sediment. For $\tau=.25$ m, however, the plots of the signals reflect the effects of the complex interactions that occur when the acoustic energy penetrates to the bottom sediment layer. These plots indicate the difficulty and ill-posedness of the inverse problem in the multi-layer setting, and why machine learning, described below, can play a valuable role in discriminating these signals based on material type or other classes.

\subsection{The data}

Backscatter signals are generated with fixed source frequency $\omega=15$ kHz, incident angle $\theta = \pi/12$ radians, and sound attenuation $\alpha= 0.5$ dB per 1 meter per 1 kHz. In the current study we chose not to distinguish attenuation based on material type. For our current purposes, we chose to isolate the small-scale effects of the seafloor by assuming that $f$ is a mean-zero function with prescribed root mean square (RMS) height and correlation length. Table \ref{table:datasets} summarizes the environmental parameters used to generate  the synthetic backscatter signals (\ref{eq:backscattered})  approximated with FEM and NMLA. The signals are generated using a random selection of the available parameters to create balanced datasets. Our rationale in designing test sets is to isolate the effects of variability in each parameter on the classification errors.

Full wave solutions are generated using the COMSOL Multiphysics{\textregistered} Acoustics Module \cite{Comsol} for the solution of the Helmholtz equations with $P^{2}-$ finite element methods on $2\text{ m } \times 2\text{ m }$ square domains, which translates to about $10^{5}$ degrees of freedom per template. For more details on sonar forward modeling using COMSOL Multiphysics{\textregistered}  see \cite{Isakson2008, Isakson2014}.  Then, NMLA  implemented in MATLAB is applied to approximate backscatter at locations $(x_{k}, z_{0})$, where the depth is $z_{0}=1$ meter above the seafloor and  $\{x_{k}\}_{k=1}^{1024}$ is a uniformly spaced horizontal discretization of the seafloor domain.

\subsection{Sediment classification}

The top part of Table \ref{table:backscattercnn} shows the accuracy of different classifiers implemented in \texttt{scikit-learn} as described in \ref{sec:baseline}. These baseline classifiers are trained on $N=20,000$ signals taken from the training data set, 20\% of which are set aside for validation. The performance of the classifiers is evaluated on 5,000 signals from each of the four different test environments with different realizations of the water-sediment interface $f$. These results indicate that the baseline machine learning classifiers do not generalize to different test sets generated from acoustic environments with slightly perturbed parameters. Based on our findings, described next, deeper neural networks can exhibit a better classification performance on the validation data and generalize better to the different test environments.

\begin{table}
\begin{center}
\footnotesize{
		\begin{tabular}{p{4cm}p{2cm}p{1cm}p{1cm}p{1cm}p{1cm}p{1cm}}
			{\bf Classifier}  &{\bf Training Time}  &\multicolumn{5}{c}{\bf Classification Accuracy  (\%)} \\
			&  & Val. &Test 1& Test 2 & Test 3 & Test 4\\
			\hline
			&  (CPU)&&&&\\
			NearestCentroid&0:00:03&63.3&66.63&64.54&53.37&64.67\\
			RandomForestClassifier&0:00:07&71.5&71.74&70.15&53.86&71.44\\
			DecisionTreeClassifier&0:00:10&55.1&55.95&54.36&43.17&53.94\\
			KNeighborsClassifier&0:00:34&55.7&59.41&57.08&46.29&57.36\\
			SVC-RBF&0:02:05&{\bf 77.3}&{\bf 76.62}&{\bf 75.76}&{\bf 56.63}&{\bf 76.19}\\
			SVC-1&0:02:18&67.4&59.44&58.14&42.98&58.36\\
			MLPClassifier&0:05:38&73.6&73.04&72.43&51.47&73.43\\
			LogisticRegression&0:45:52&67.7&57.38&55.74&41.54&56.36\\
			\hline
			&  (GPU)&&&&\\
			AlexNet-1D &00:05:30 &88.95& {\bf 89.72}   &{\bf 88.46}   &{\bf 58.19}&   {\bf 88.78}\\
			CNN-3 &	00:08:28  &91.61&43.45   &42.49  &37.18&   42.60\\
			CNN-4	& 00:10:47  &{\bf 92.41}& 24.34   &24.10   &24.08&   24.12\\
			ResNet50-1D	 & 01:00:20 &89.62&26.16   &25.08   &25.09&   25.02\\
			GoogleNet-1D	&07:30:01 & 86.03& 86.14   &{ 86.61}   &57.95&   85.88\\
			VGG19-1D	 &31:37:46 &90.31&26.19   &25.12   &25.15&   25.04\\
			
		\end{tabular}
		}
		\end{center}
	\caption{Training time and accuracy of different shallow machine learning tools (top) and DNNs (bottom) applied to a validation set (no added noise) and four test sets (noise added with SNR = 20).}\label{table:backscattercnn}
\end{table}

\subsubsection{Implementation of CNNs}

\begin{figure}
	\includegraphics[width=0.99\textwidth]{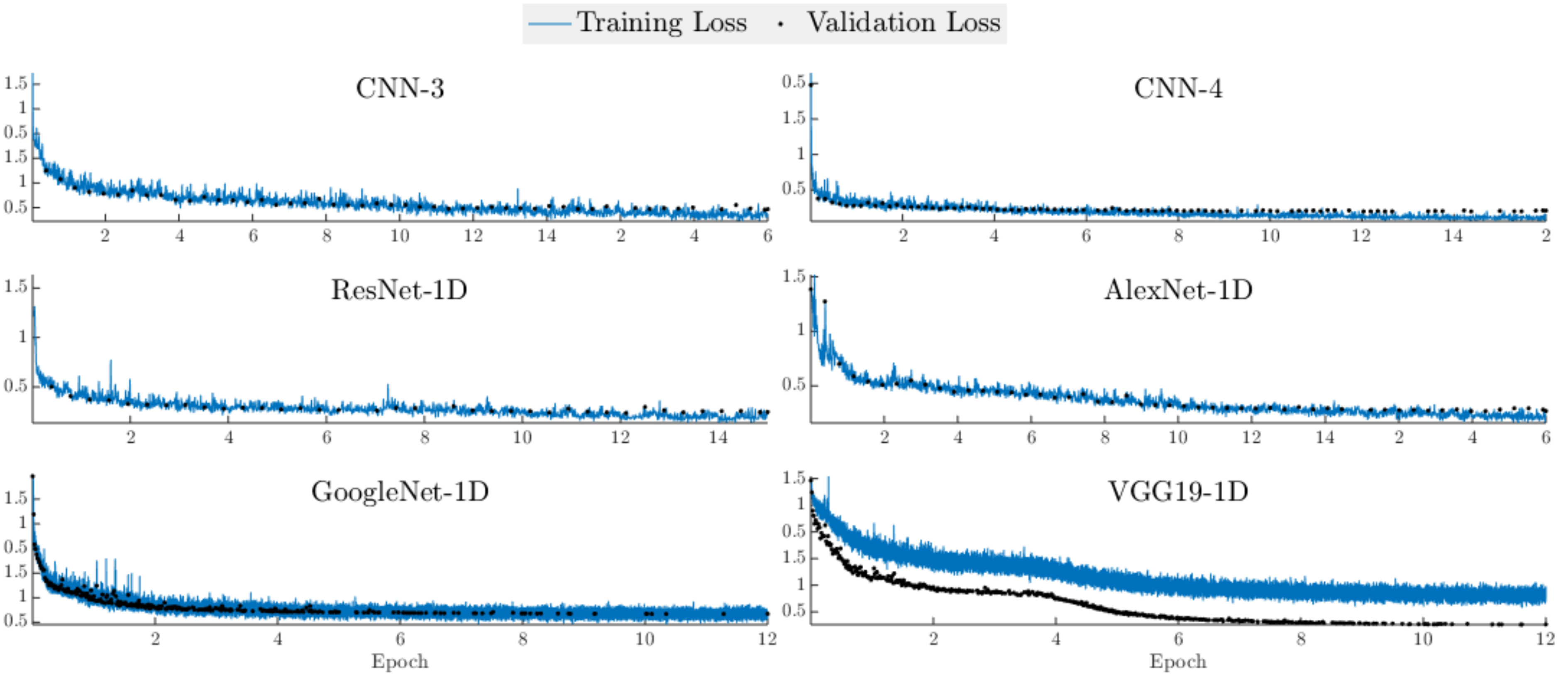}
			\caption{Cross entropy loss vs. training epoch for six DNN classifiers.}\label{fig:training}
	\end{figure}

Initially, we constructed two smaller convolutional neural networks,  `CNN-3' and `CNN-4', having three and four convolutional layers, respectively. In both cases, each convolutional layer is followed by a batch normalization layer, ReLU activation functions, and a max pooling layer, and the filter sizes vary from 16$\times$1 to 256$\times$1. We also considered the following well-known deeper CNNs by adapting them to the 1D setting: 	AlexNet \cite{krizhevsky2012imagenet} (8 convolutional layers), GoogleNet \cite{szegedy2015going} (22 convolutional layers), ResNet50 \cite{he2016deep} (50 convolutional layers), and VGG-19
\cite{simonyan2014very} (19 convolutional layers). MATLAB code and the trained models are publicly available\footnote{\url{https://github.com/cf87/Seabed-Classification-2020}}.

We use the Deep Learning toolbox in MATLAB to train convolutional neural nets (CNNs) using $N=80,000$ signals taken from the training data set, 20\% of which are set aside for hold-out validation. The performance is evaluated using 5,000 signals from each of the four different test sets. The labels of the test signals correspond to the closest match, in terms of absolute distance, between the sound speed of the top layer and the four sound speed classes used in the training dataset.

Learnable parameters in each CNN are determined using the ADAM method for stochastic optimization \cite{kingma2014adam}. Training options for the ADAM optimizer were chosen based on trial and error. We chose a mini-batch size of 500; the learning rate schedule varied between $1\mathrm{e}{-4}$ and $1\mathrm{e}{-3}$, with a 50\% drop in the learning rate scheduled periodically (every 10-50 epochs) during the training; the training data is shuffled before every epoch; the validation data is shuffled before each network validation;  the gradient threshold was set to 1; and the $L_{2}$ regularization factor was set to $1\mathrm{e}{-4}$. The termination criteria was determined to ensure a decreasing validation loss. For the deeper CNNs, the training was performed on one or two NVIDIA Tesla P100 16GB GPUs in parallel. Figure \ref{fig:training} shows plots of the cross-entropy training loss and validation loss for six different CNN architectures. The training times and test accuracy (discussed next) for each classifier are listed in Table \ref{table:backscattercnn}.

\subsubsection{Results}

\begin{figure}
\begin{center}
	\includegraphics[height=0.3\textwidth]{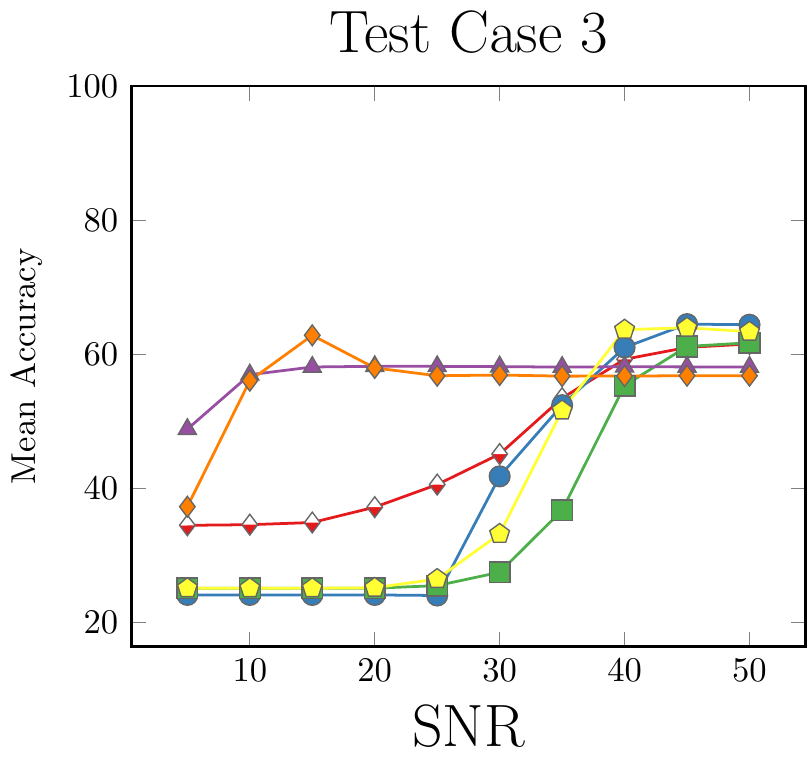}
	 \includegraphics[height=0.3\textwidth]{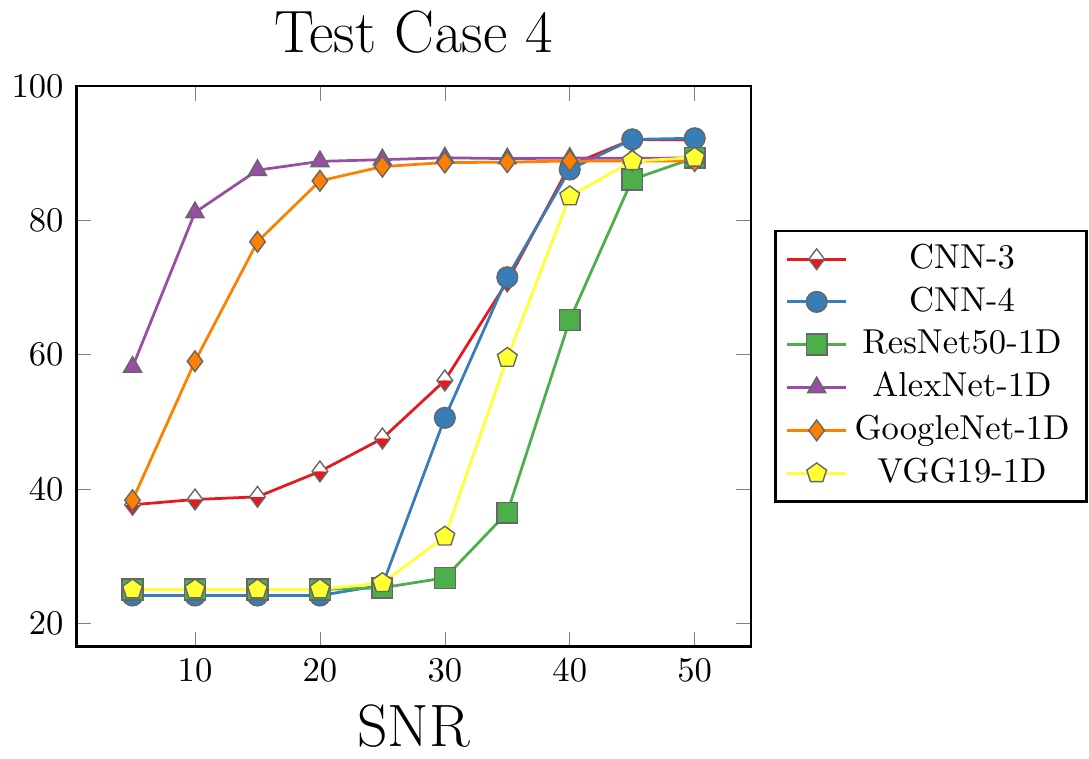}
	 \end{center}
	\caption{ Classification performance of different trained CNNs on the (left) Test 3 and (right) Test 4 environments.}
	\label{fig:allDNNsTest4}
\end{figure}

Figure \ref{fig:allDNNsTest4} (left) and (right) contain plots of the mean accuracy of different CNNs used to classify test data with varying SNR. The mean is taken over four realizations of noise added to each of the 5,000 signals, resulting in 20,000 predicted labels. Out of the classifiers that we tested, the CNNs that showed the most robustness to noise were the adapted AlexNet-1D, GoogleNet-1D, and to a lesser extent, CNN-3. This performance was also seen for the other test environments not shown.

\begin{figure}[t]
\begin{center}
	\includegraphics[height=0.34\textwidth]{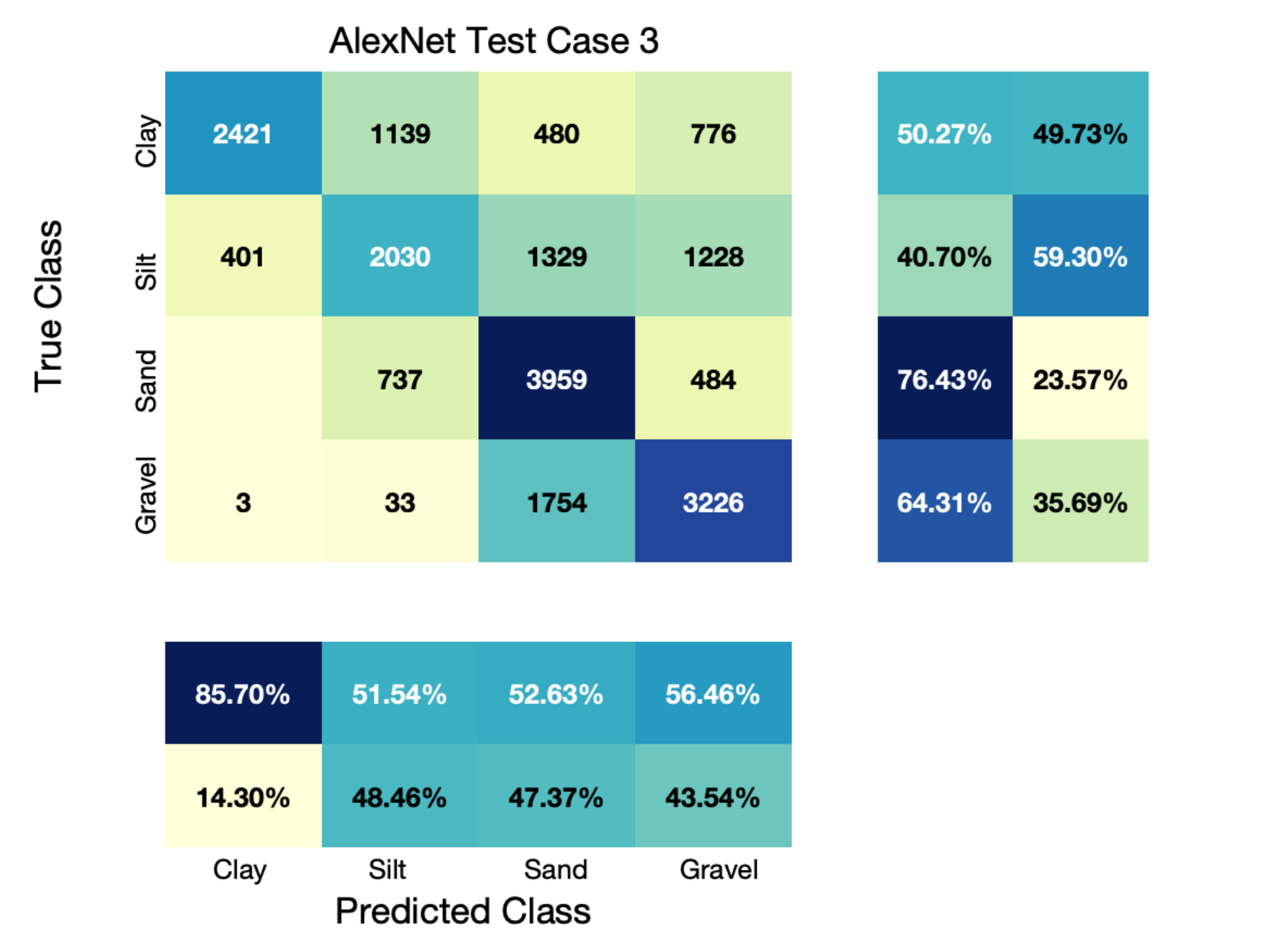}
	\includegraphics[height=0.34\textwidth]{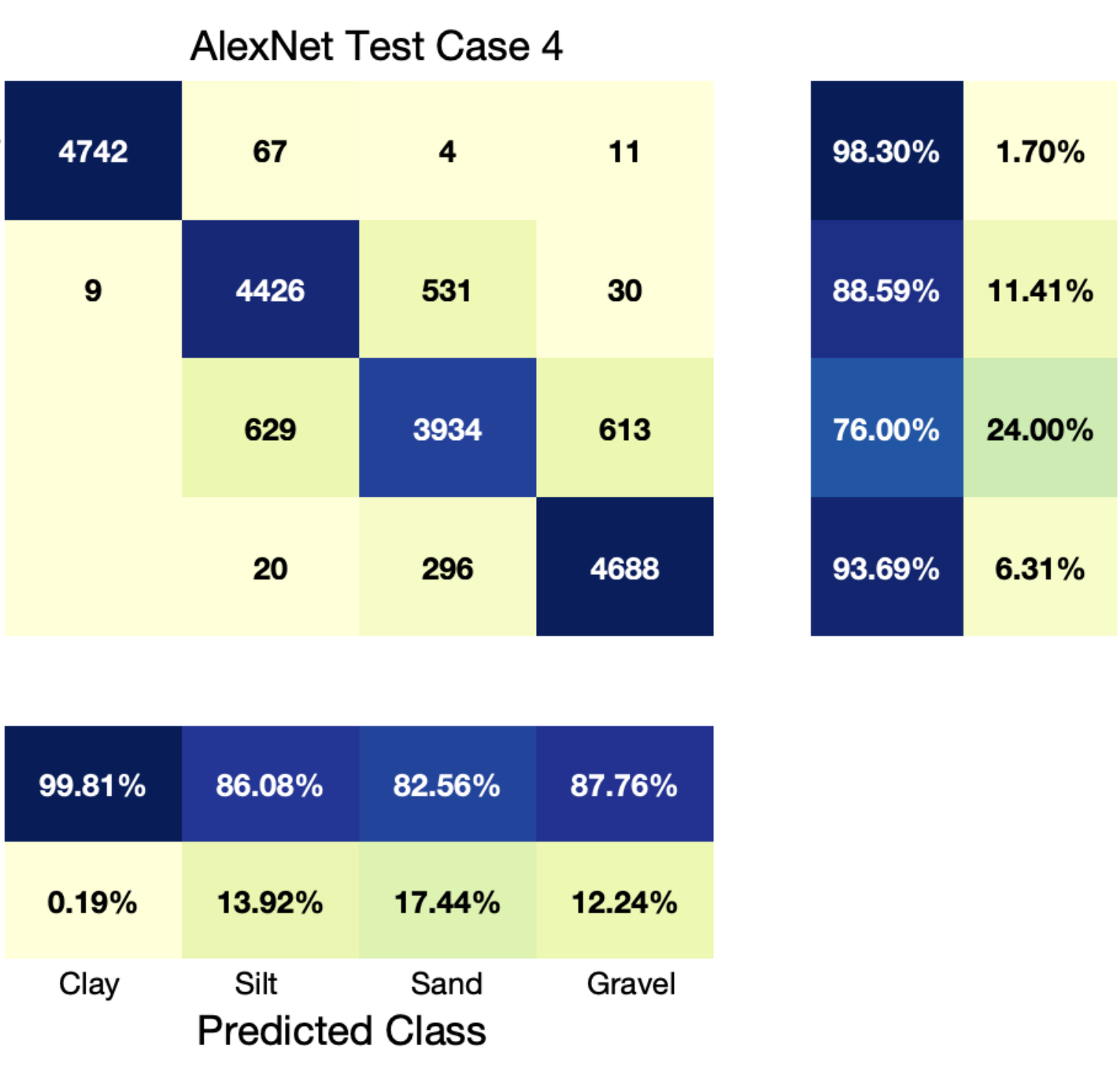}
	\end{center}
	\caption{ Confusion matrices for adapted AlexNet-1D for noisy test data (SNR = 20 dB) on the (left) Test 3 and (right) Test 4 environments.}
	\label{fig:conf}
\end{figure}

One criticism that deep learning techniques receive is their lack of interpretability,  in particular when compared with classifiers like SVM and KNN. In scenarios like seabed classification, physical modeling offers the advantage of being able to test many different environments to gain insight on the generalization of classifiers. The four test environments we developed allow us to understand the sensitivity of physical parameters to the machine learning outcomes. The charts in Figure \ref{fig:conf} are confusion matrices for the best classifier, the AlexNet-1D, corresponding to noisy test data with an SNR of 20. There was significant misclassification of signals in Test 3 corresponding to clay and silt responses. Based on the test environments, these results indicate that the layer thickness can have significant influence on misclassification rates. This is consistent with physical intuition, and indicates the need to resolve the layer thickness with high accuracy to obtain better classification results.  On the other hand, the classifier performance in Test 4 shows improved classification.

Although the adapted GoogleNet-1D had comparable accuracy and robustness to noisy data, the training times, listed in bottom part of Table \ref{table:backscattercnn},  indicate the superiority of the adapted AlexNet-1D (under 11 minutes) over GoogleNet-1D (over 7 hours). Although the smaller, custom CNNs have low training times and performed well on test sets with low SNR, they were not shown to be robust to added noise. Even at low SNR, AlexNet-1D outperforms the best baseline machine learning classifiers, however, in the case of noisy data, the baseline classifiers performed much better overall than the custom CNNs, ResNet50-1D, and VGG19-1D. Interestingly, the SVC performed the best on the Test 3 dataset corresponding to an environment with perturbed layer thickness, which was the most challenging test case.

An interesting outcome of these numerical experiments is that some of the deep learning classifiers that perform well on the validation set are vulnerable to noise and small perturbations in the geoacoustic properties of the test environments. Both AlexNet-1D and GoogleNet-1D demonstrated good generalization behavior whereas
ResNet-1D, VGG19-1D and CNN-4 did not generalize at all. In fact, the poorest performing classifier makes predictions that have about the same accuracy as a random guess. Understanding in what context deep learning classifiers are good at generalizing is currently an
open question within machine learning (see for instance \cite{recht2018cifar}).

\subsection{Further Discussion}

\begin{figure}[t!]
	\includegraphics[width=0.37\textwidth]{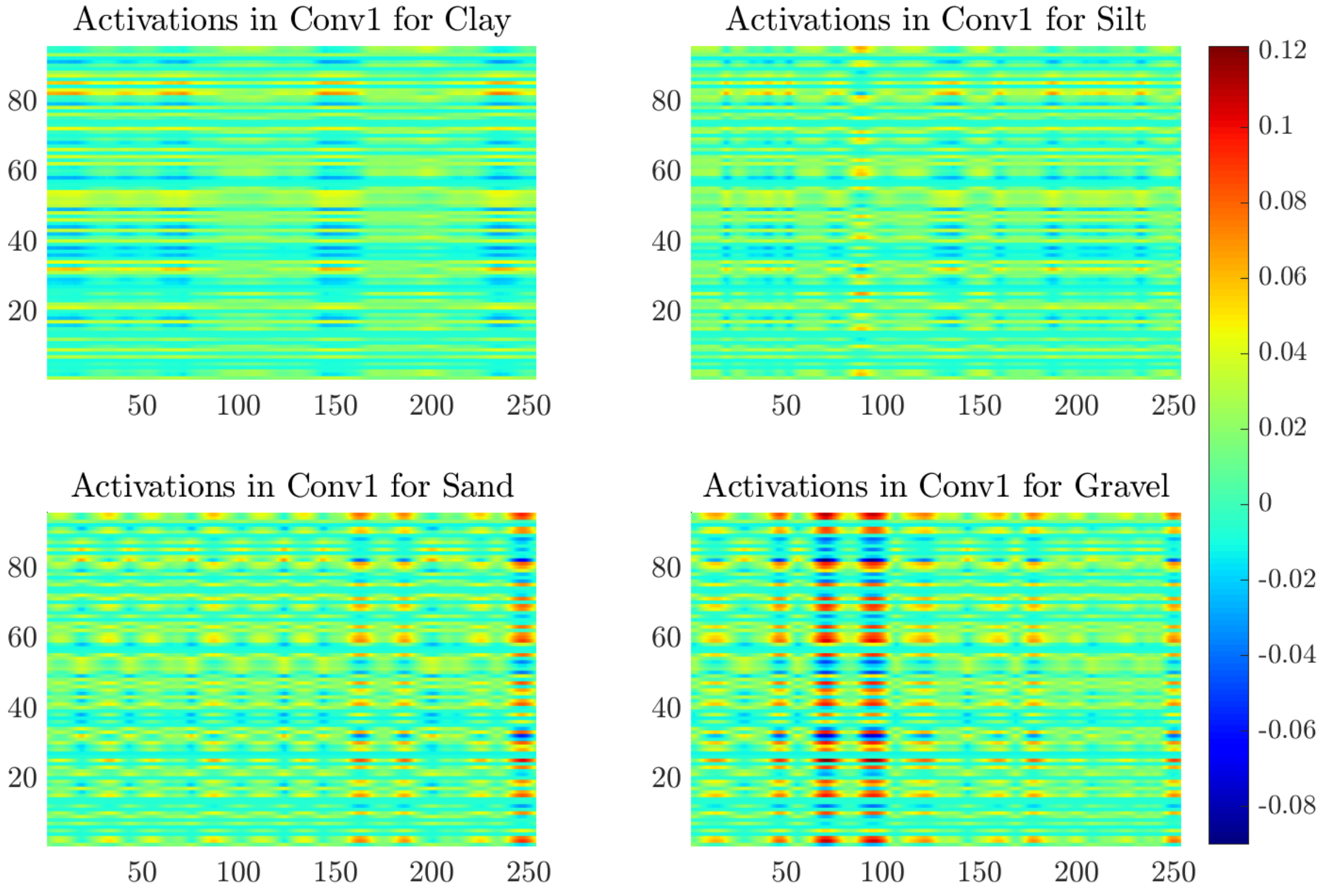}
	\includegraphics[width=0.62\textwidth]{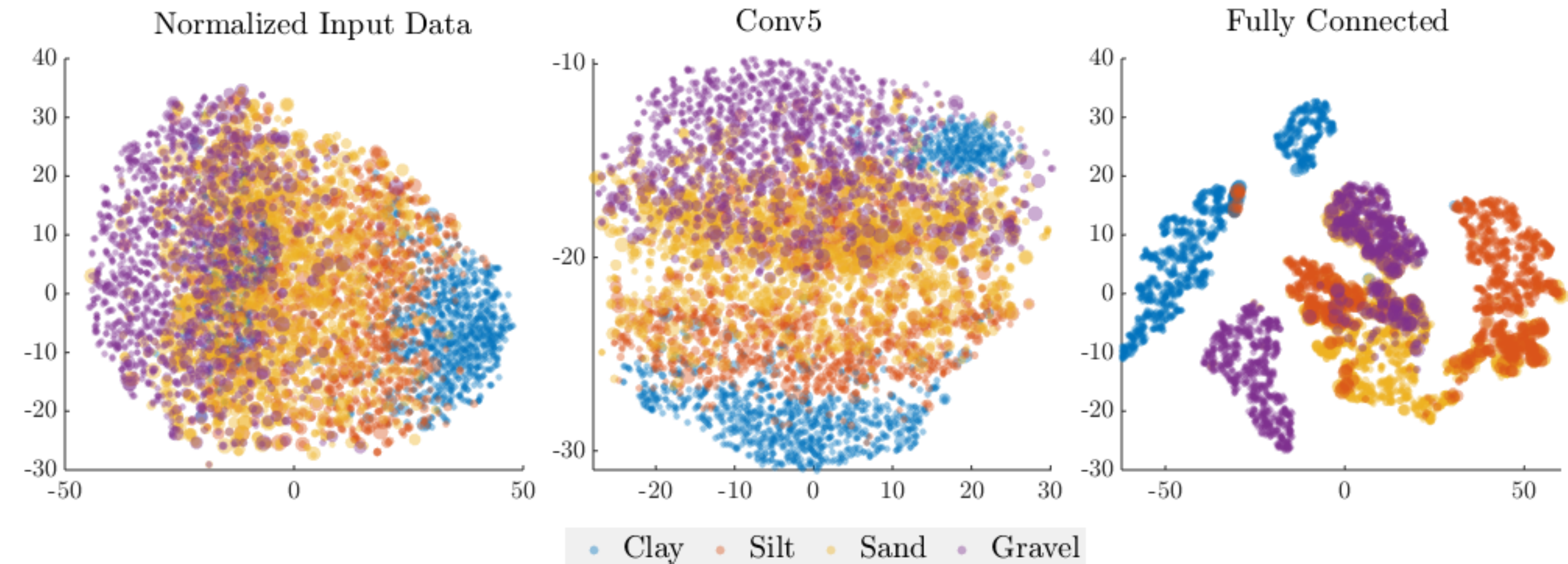}
	
	\caption{(left)  Activations in the `Conv1' layer of AlexNet-1D. (right) T-SNE visualization of three different layers of AlexNet-1D.}\label{fig:testactivations}
\end{figure}

 One way to visualize the behavior of deep neural networks by plotting the activation patterns of different layers for a given input.  Figure \ref{fig:testactivations} (left) contains plots of the activations of the first convolutional layer in the trained AlexNet-1D for four different input test signals. The pixel intensities signify which channels are activated. This plot gives a sense of which filters in each layer are activated by different inputs, for example, the filters in the 50-100 range are activated by input signals from gravel much more than the other three materials. Furthermore, these plots highlight the ease of separating sand and gravel signals, which strongly activate filters in this layer, as opposed to silt and clay, which only have weak activations.

Viewing activations in this way across many signals can be challenging, since the set of activations in each layer corresponds to a high dimensional vector. One way to approach the task is the t-Distributed Stochastic Neighbor Embedding (t-SNE), \cite{maaten2008visualizing}, which constructs a probability distribution defined on pairs of activation vectors, where the probability of sampling two vectors increases as the distance between the vectors decreases. Typically, this distribution is modeled with a Gaussian kernel. Then, gradient descent is used to find a map between the high dimensional distribution and a $t-$ distribution, defined on a low dimensional space, that minimizes the Kullback$-$Leibler divergence.  The t-SNE plots in Figure \ref{fig:testactivations} (right) correspond to the layer activations for all of the signals in the Test 4 dataset corresponding to the first max pooling layer, final convolutional layer, and the softmax layer that produces a probability distribution on the set of possible classes. This visualization suggests that in the initial layer and fifth convolutional layer, the activations corresponding to different class boundaries have overlaps, however, in the final fully connected layer, the class boundaries are more separated. This suggests that the full architecture of AlexNet-1D is really needed to produce accurate classifications.

Another technique used to interpret the deep net classifier is 'GradMap'
\cite{selvaraju2017grad}. The method involves using the gradient of the loss
function to identify relevant patterns within the signal. The gradient of
the loss function can also be used to identify adversarial attacks to
the classifier \cite{goodfellow2014explaining}. Such analyses are
beyond the scope of this work.

\section{Conclusions}

We employ a model-based approach for designing training and test datasets of acoustic templates for capturing the relevant physics of representative patches of seafloor. At low frequencies, this is accomplished with normal mode propagation, and at higher frequencies, local modeling on smaller computational domains enables fast, parallelizable simulations. An underlying assumption is spatial stationarity of the seafloor, which is reasonable in situations where roughness statistics are similar over an area larger than the ensonified area. In this study we performed sediment classification in a two layer seafloor, varying both geoacoustic parameters (sound speed, density) and geometric parameters (roughness, thickness) in the training data. 

In low-frequency models, standard machine learning classifiers, such as logistic regression models and support vector machines, outperformed traditional matched-field processing techniques, especially when the test data had a low signal-to-noise ratio. Confusion matrices corresponding to the models indicate that certain classes have a higher likelihood of misclassification, namely silt and clay. On the other hand, predictions of gravel and sand are more likely to be correct. For backscatter data, standard machine learning classifiers have poor accuracy and do not generalize well to other test environments. Some of the deep learning classifiers, namely AlexNet and GoogleNet, adapted to 1D signals, are more costly to train but produce higher accuracy classification and better generalization. These results also indicated that the layer thickness can have significant influence on misclassification rates. Further investigation must place an emphasis on resolving the layer thickness with high accuracy.

Our results indicate the promise of machine learning and deep learning for the difficult problem of geoacoustic classification. The results from our simulations highlight the need to test models for a broad spectrum of environments to ensure generalization. Producing a well-performing deep learning model requires thorough experimental design. There are several other directions that can be explored with this framework, for example, finding elastic properties of sediments and incorporating the influence of the material type on the statistical properties of the seafloor roughness.

\section*{Acknowledgements}
The authors are grateful to Kyunghyun Cho (CDS, NYU) for valuable discussions. The research of CF is supported by NSF DMS 1720306.  The research of ZHM is supported by ONR N000142012029, N000141612485, and N000141812125. SV is partly supported by NSF DMS 1913134, EOARD FA9550-18-1-7007 and the Simons Algorithms and Geometry (A\&G) Think Tank.

\end{document}